\documentclass[amsmath,amssymb,aps,pre,showpacs, twocolumn]{revtex4}

\usepackage[T1]{fontenc}
\usepackage{graphicx}
\usepackage{epstopdf}
\usepackage[normalem]{ulem}

\graphicspath{{images/}{plots/}}

\newcommand{\change}[1]{{\bf #1}}

\usepackage{mathptmx}
\DeclareSymbolFont{symbols}{OMS}{cmsy}{m}{n}
\DeclareSymbolFont{largesymbols}{OMX}{cmex}{m}{n}
\usepackage[]{hyperref}
\hypersetup{
  colorlinks,
  linkcolor={blue},
  citecolor={blue},
  urlcolor={blue}
}

\begin{document}

\title{Exact Ground States of the Kaya-Berker Model}
\author{Sebastian \surname{von Ohr}}
 \email{sebastian.von.ohr@gmail.com}
\author{Alexander K. Hartmann}
\affiliation{Institute of Physics, Carl von Ossietzky University, 26111 Oldenburg, Germany}
\date{\today}

\begin{abstract}
Here we study the two-dimensional 
Kaya-Berker model, with a site occupancy $p$ of one 
sublattice,  by using a polynomial-time  exact ground-state
algorithm. Thus, we were able to obtain $T=0$ results in exact equilibrium
for rather large system sizes up to \change{$777^2$} lattice sites. 
We obtained sublattice magnetization
and the corresponding Binder parameter. We found a critical point 
$p_c=0.6423(3)$ beyond which the sublattice magnetization vanishes.
 This is clearly smaller than previous results which were
obtained by using non-exact approaches for much smaller systems.
We also created for each realization minimum-energy
domain walls from two ground-state calculations for
periodic and anti-periodic boundary conditions, respectively. The analysis
of the mean and the variance of the domain-wall distribution shows that
there is no thermodynamic stable spin-glass phase, in contrast to previous
claims about this model.
\end{abstract}


\maketitle

\section{Introduction} 

Compared to regular or pure systems, magnetic systems with
quenched disorder, like spin glasses and random-field systems 
\cite{young1998}, exhibit many peculiar properties.
 Their complex low-temperature
behavior is still not fully understood, even for two-dimensional
systems. As analytical solutions are not available, computer simulation
studies \cite{practical_guide2015} are often performed. 
With respect to Markov-chain Monte Carlo simulations
\cite{newman1999}, 
one of the difficulty is their slow glassy dynamic, resulting
in very long equilibration times. Other approaches to study
spin-glasses involve finding and characterizing ground states
\cite{opt-phys2001,opt-phys2004}. In
three or more dimensions, only algorithms with exponential running time are
known, but in two dimensions polynomial time algorithms are
available. One of the interesting differences between 2D and 3D
spin-glasses is that so far all 2D models with short or finite-range
interactions show a transition temperature $T_\text c =0$ 
\cite{mcmillan1984,bray1984,saul1993,rieger1996,houdayer2001,%
hartmann2001,carter2002,hartmann2003}. 
At all finite temperatures the spin-glass phase vanishes for 2D
models. In contrast the 3D models show a transition temperature above
zero. This spawned the search for 2D models with a finite critical
temperature. One such candidate is the Kaya-Berker model \cite{kaya2000},
which was claimed to exhibit a spin-glass like phase for non-zero
temperatures, i.e., a phase transition at a finite temperature. 
This previous claim was based on numerical studies of rather small systems
with non-exact algorithms. Here, we will
present results for this model which we obtained by using exact and 
fast ground-state algorithms. This allowed us to study in exact equilibrium
rather large systems exhibiting more than $10^5$ spins. Our
result strongly suggest that in contrast to previous claims, the model does
\emph{not} exhibit an ordered low-temperature phase.

The manuscript is organized as follows: We will first introduce the model 
along with suitable measurable quantities and review
previous results. Next, we outline the algorithm we have used to obtain
exact ground states, and, by changing the boundary conditions, 
to obtain domain-wall (DW) energies. In the main part, we will 
present our results, followed by our conclusions.

\section{Model}
\begin{figure}
 \includegraphics[width=0.95\columnwidth,trim={0 60px 0 60px},clip]
{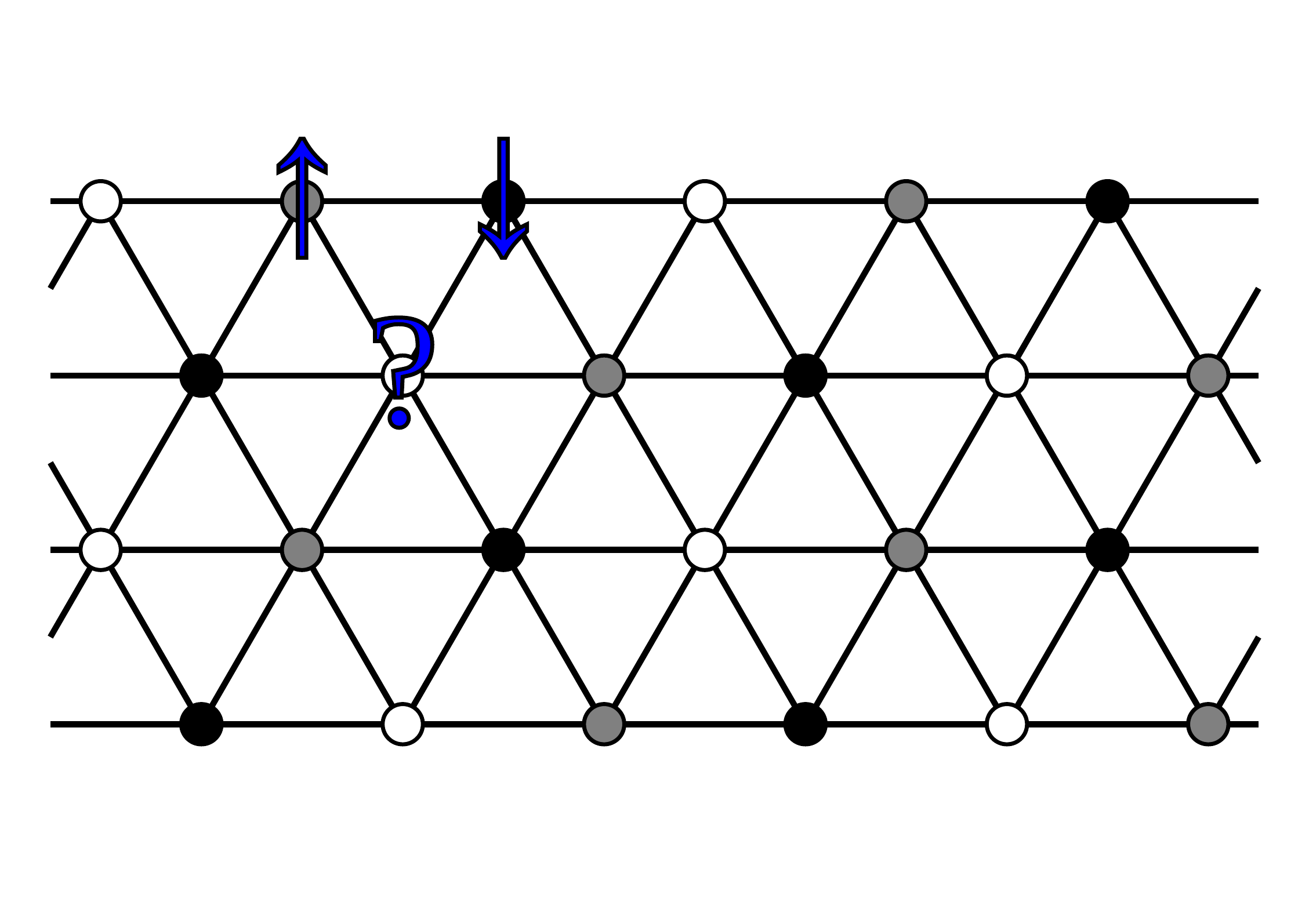}
 \caption{(Color online) $6 \times 4$ triangular lattice showing the subdivision into the 
three sublattices, as indicated by the three colors of the nodes (white, 
gray, black). Every triangle is frustrated, since not all bonds can be 
satisfied.}
 \label{fig:tri_lattice}
\end{figure}

The Kaya-Berker Model \cite{kaya2000} is a variation of the Ising model on a two-dimensional triangular lattice with $N = L_x \times L_y$ spins. The spins $s_i$ take the values $\pm 1$ and all bonds are antiferromagnetic. It's Hamiltonian is given by
\begin{equation}
 \mathcal H = -J \sum_{\langle i, j\rangle} \epsilon_i s_i \epsilon_j s_j \,,
\end{equation}
with $J < 0$ and $\langle i, j\rangle$ indicating a sum over all 
nearest-neighbor pairs. The model allows for dilution, which is
described by the quenched disorder variables $\epsilon_i\in\{0,1\}$.
Every spin is located on one of three sublattices, such that every 
spin has only neighbors in the two other sublattices. 
Fig.~\ref{fig:tri_lattice} shows the triangular lattice and 
subdivision into three sublattices. Here, one of the sublattices 
is diluted and 
only a fraction $p$ of spin sites is occupied ($\epsilon_i=1$),
while a fraction $1-p$ of sites is not occupied by a spin ($\epsilon_i=0$).

In the fully occupied $p=1$ case, every triangle of spins is frustrated. 
This special configuration was solved exactly 
\cite{wannier1950, houtappel1950}, with the result that the 
system is disordered at all temperatures. Ground states are 
characterized by exactly one-third of unsatisfied bonds. 
While there are some configurations that are ordered, e.g., 
alternating row of all up-spins with row of all down-spins, 
no energy advantage is obtained for long-range order. Because of
entropic dominance, i.e., the exponential dominance of these
non-ordered ground state configurations, no long-range order occurs

For the diluted $p=0$ case, one obtains a honeycomb lattice, where the frustration is fully relieved. The ground state is ordered and spins are aligned antiparallel with all of their neighbors. Within the two remaining sublattices spins in the same sublattices are aligned in the same direction.

By choosing an intermediate value of $p$, the number of 
frustrated plaquettes can be varied. The behavior of the system is 
complicated and allows for interesting behavior. These intermediate 
values of $p$ result in a ground state with zero magnetization on 
the diluted lattice and roughly equal but opposite magnetization on 
the two undiluted lattices. The order parameter is therefore defined as 
the per-lattice magnetization
\begin{equation} \label{eq:magnetization}
 m_\alpha = \frac{1}{N_\alpha} \sum_{i \in \alpha} \epsilon_i s_i \,,
\end{equation}
with $\alpha=a,b,c$ denoting one of the three sublattices
and $N_\alpha=\sum_{i \in \alpha} \epsilon_i$ is the number of spins
of sublattice $\alpha$. 
The diluted lattice will be lattice $a$.

A model similar to the Kaya-Berker model with uniform dilution in all 
sublattices \cite{grest1979, blackman1981, anderico1982, tang2010} 
was studied earlier. While the first study observed spin-glass behavior, 
all later publications argue against a spin-glass phase. However, 
they found a large but finite correlation between spins, which 
could be mistaken for long-range order in small systems. In 
2000 H.~Kaya and A.~N.~Berker devised the aforementioned model,
which notably differs from the older model 
by restricting the dilution to
one sublattice \cite{kaya2000}. The authors
studied it  using hard-spin mean field (HSMF) 
theory. For HSMF, each spin $s_i$ does not only interact with
its neighbours $s_j$ through their mean spin values $m_j$
as in standard mean-field theory. Instead, the
self-consistent equation for the site-dependent 
mean values $m_j$ involves a sum over all possible
$2^n$ configurations of the $n$ neighbours such that each spin
orientation $s_j=\pm 1$ occurs with a probability which is 
compatible with its mean value $m_j$. As a further approximation,
in Ref.~\cite{kaya2000} the disorder average is performed and 
the site dependent mean values are replaced
by their sublattice mean values, 
resulting in three coupled equations.
For the sublattice spin-glass order parameter (which involves again
the site-dependent mean values)
\begin{equation}
q_\alpha = \left[\frac 1 
{N_\alpha}  \sum_{i \in \alpha} (m_i-m_\alpha )^2 
\right]^{1/2}\,,
\end{equation}
they found nonzero values at 
finite temperatures for occupancy $p < 0.958$. However, their 
study involves only small system with sizes up to $30 \times 30$ and 
features no finite-size scaling analysis. Other studies analyzed the 
Kaya-Berker model using Monte Carlo (MC) simulations 
\cite{robinson2003, robinson2011}, effective-field theory (EFT) 
\cite{zukovic2012} and a modified pair-approximation (PA) method 
\cite{balcerzak2014}. The EFT approach is based on a cluster 
approximation with clusters comprised of only a single spin and 
interactions with their nearest neighbors \cite{zukovic2010}, 
which is quite similar to HSMF. The PA method is based on the 
cumulant expansion of the entropy. So far, none of the studies 
found conclusive evidence in favor or against the spin-glass phase at 
finite temperatures. Note that 
using MC simulations the system certainly appears to behave
like a glassy system in the accessible system sizes and timescales. 
Here simulations at different and large enough system sizes and
proof of proper equilibration would be needed, which is difficult 
and requires a huge numerical effort for glassy systems.

This 
work aims to settle this open question by using an exact ground state 
algorithm which allowed us to investigate large systems in
equilibrium. Studying exact ground states allows calculating domain 
wall energies, which are often used 
\cite{bray1984,mcmillan1984,hartmann1999,hartmann2001,amoruso2003,roma2007} 
to verify the stability of a phase at finite temperatures. In the
following section we explain our numerical approaches.

\section{Algorithm}
\begin{figure*}
 \begin{minipage}[t]{0.24\textwidth}
  \includegraphics[width=\textwidth]{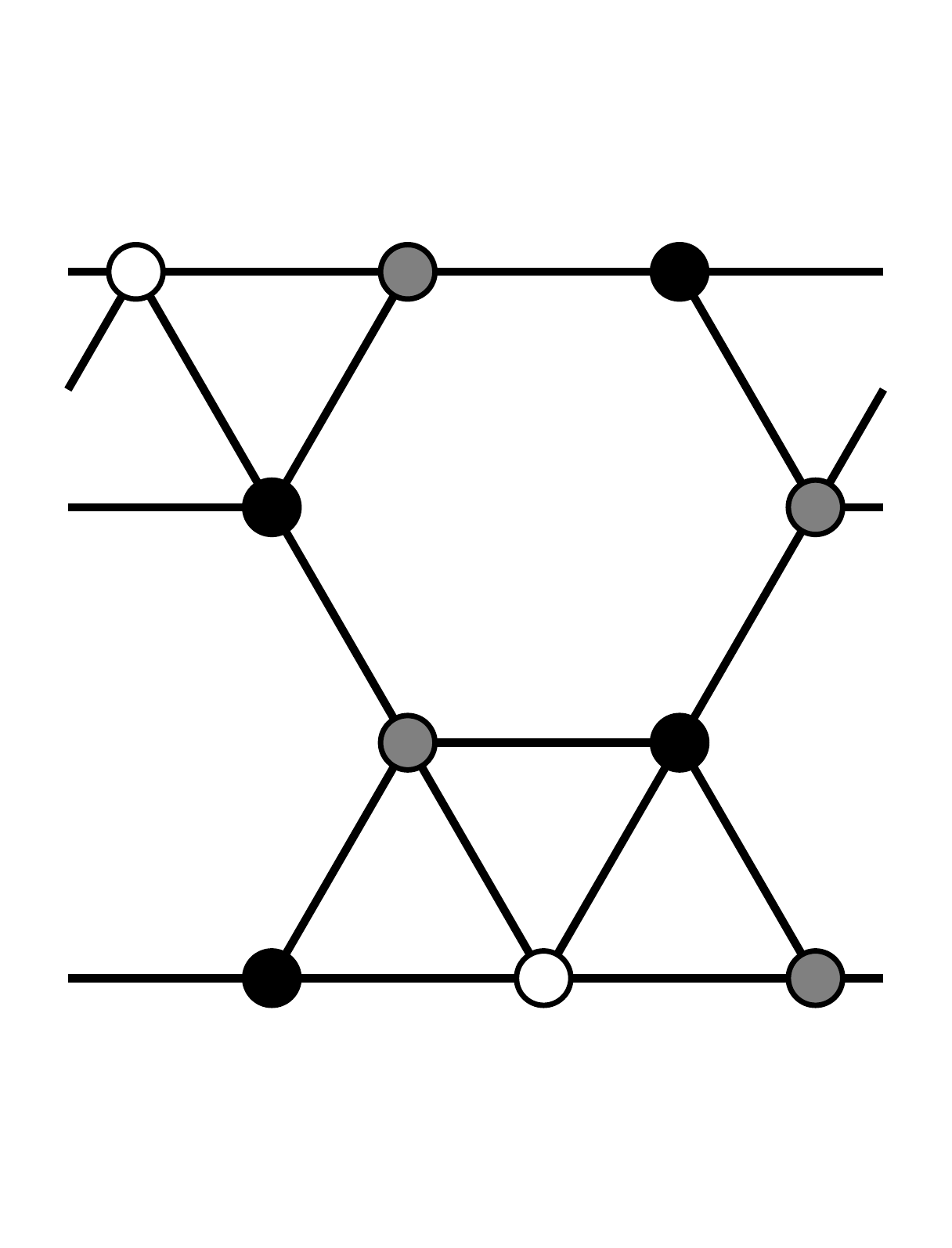}
 \end{minipage}
 \begin{minipage}[t]{0.24\textwidth}
  \includegraphics[width=\textwidth]{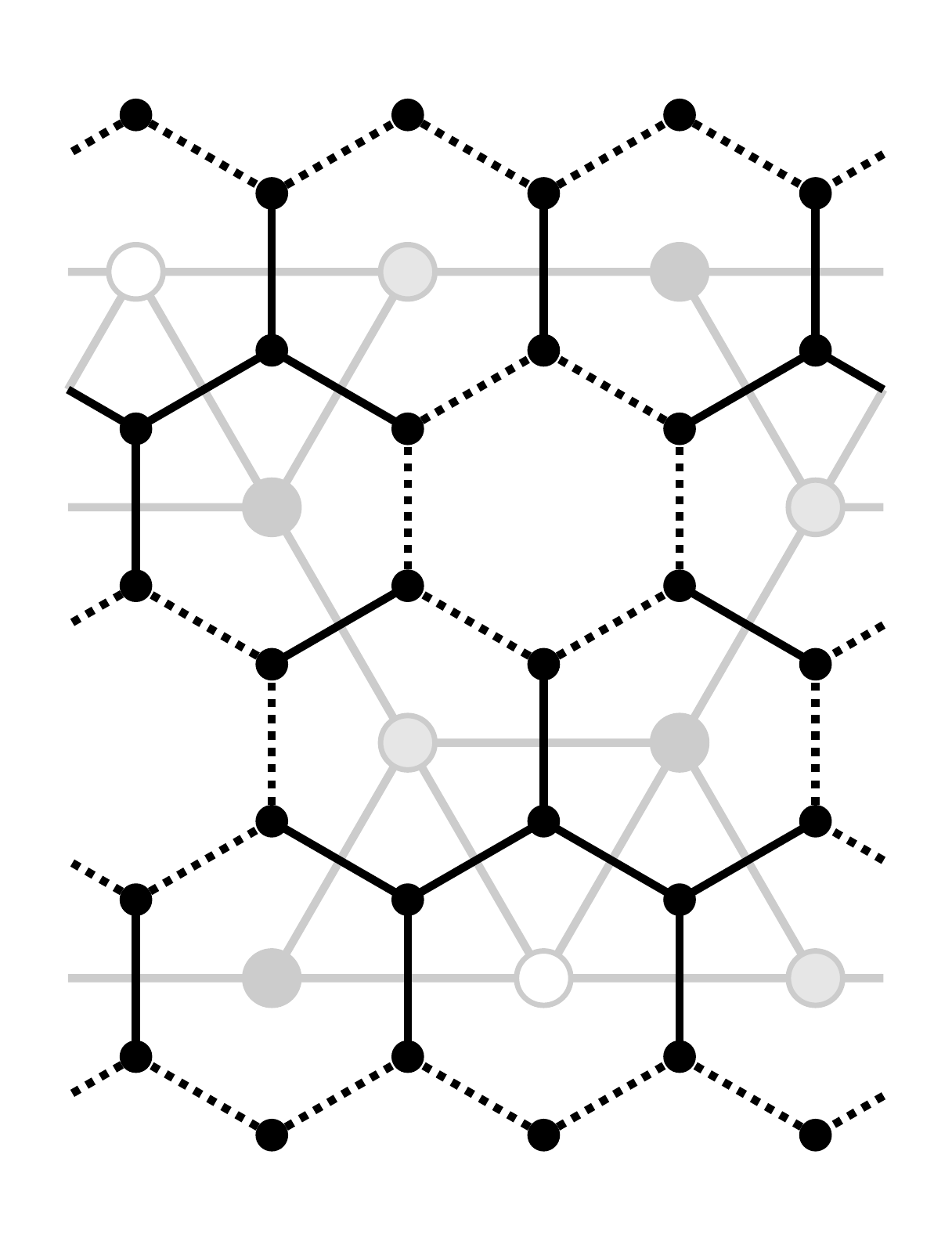}
 \end{minipage}
 \begin{minipage}[t]{0.24\textwidth}
  \includegraphics[width=\textwidth]{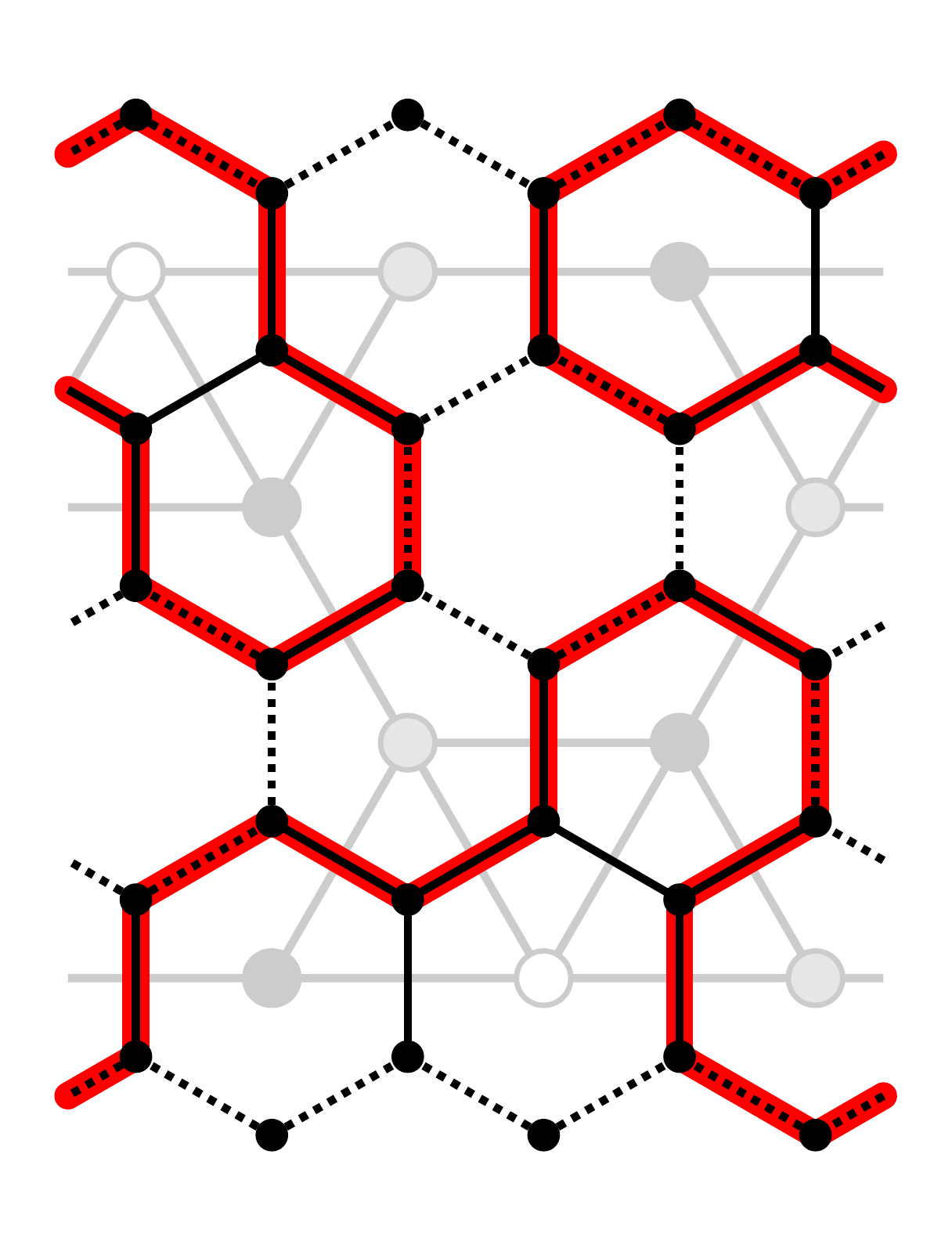}
 \end{minipage}
 \begin{minipage}[t]{0.24\textwidth}
  \includegraphics[width=\textwidth]{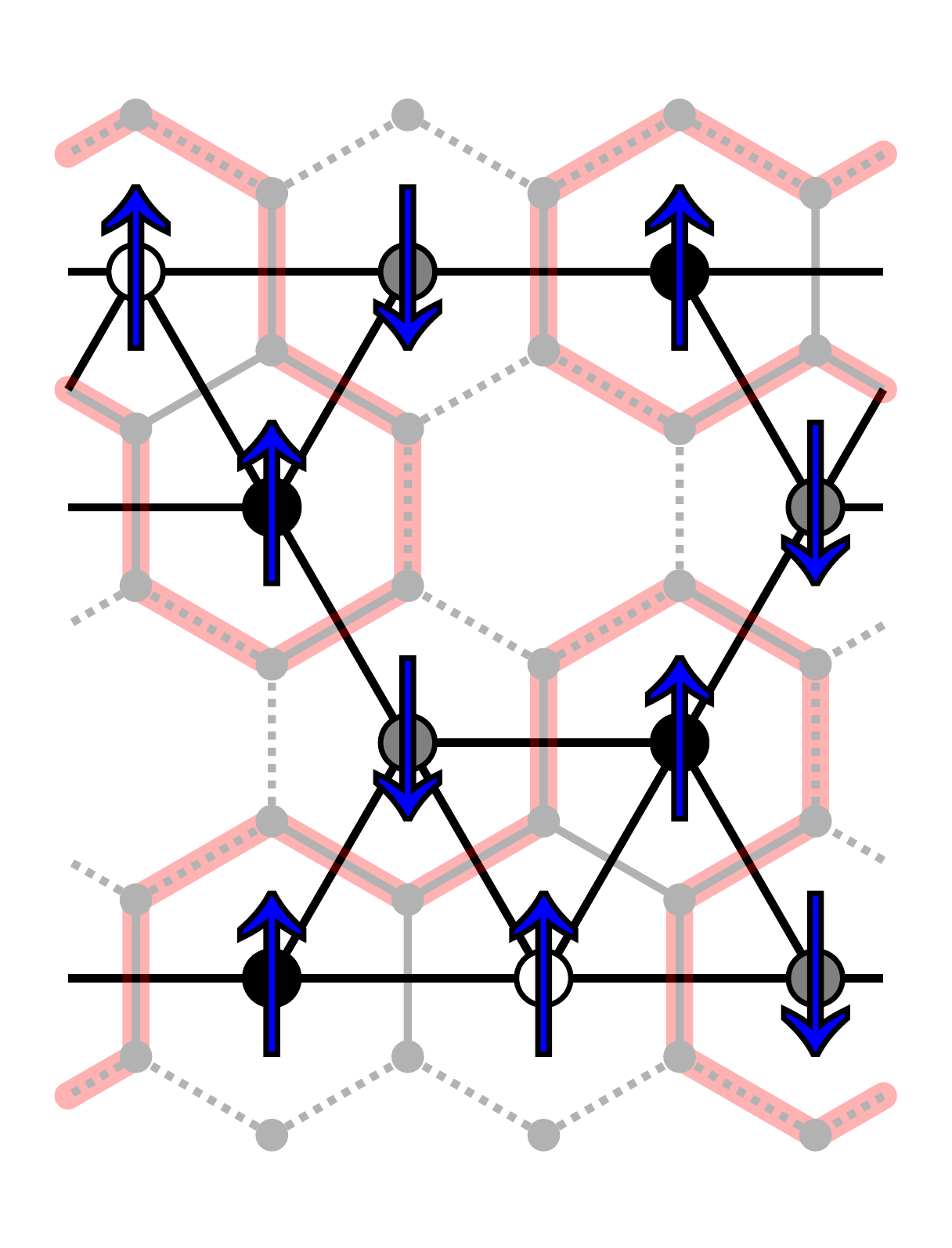}
 \end{minipage}
 \caption{(Color online) Illustration of the steps involved in the ground state calculation. The sample shows a $3 \times 4$ system with periodic boundary conditions in the horizontal direction.
 (a) Initial system with $p=0.5$ occupation on the white sublattice.
 (b) Construction of the dual graph of the original undiluted system with additional vertices and edges at the top and bottom. Edges that do not correspond to edges in the diluted system carry zero weight and are marked using stroked lines.
 (c) Set of minimal weighted loops on the dual graph. These paths separate clusters of aligned spins.
 (d) Ground state constructed by setting the top left spin to \textit{up} and assigning all remaining spins using the clusters determined earlier.}
 \label{fig:algo_steps}
\end{figure*}

The ground state algorithm is taken from Ref.~\cite{melchert2011} 
where it's used to calculate ground states of the 2D random bond Ising model on a planar triangular lattice. Here we present only a short summary of the algorithm, visualized in Fig.~\ref{fig:algo_steps}.

We start with a specific realization of our system, that we 
want to calculate a GS for. Lattice sites and bonds 
are interpreted as 
the nodes and edges of a a weighted undirected graph $G$. 
The weights are given by the strength of the bonds. For this example, 
all weights are set to $-1$, but the algorithm is suitable for arbitrary 
weights. Here we consider periodic boundary conditions
in horizontal ($x$) direction and open boundary conditions in vertical
direction.

Next, we construct an auxiliary graph $G'$, which is the dual 
graph of the undiluted triangular lattice, but with two additional 
rows of each $2 L_x$ vertices at the top and bottom, respectively.
The resulting graph contains a total of 
$(2 L_x) \times (L_y + 1)$  vertices and has the structure of a 
honeycomb lattice. Since the $2L_x\times(L_y-1)$ faces 
in $G$ are separated by single edges, there is a 
corresponding dual edge in $G'$ for every edge in $G$. Each dual edge
basically ``crosses'' the corresponding edge from $G$. These 
edges carry the same weight as the corresponding edges in $G$.
Note that the dilution is modelled by making the bonds site
dependent $J\,\to\,J_{i,j}\in\{0,J\}$ and setting the weight to zero of edges
which are adjacent to non-occupied sites. Also
the additional edges in $G'$, i.e., all the edges at the top and bottom 
which do not correspond to an edge in $G$,
are assigned zero weight.

In the next step, a set of closed non-intersecting loops with minimum 
edge weight are calculated. Calculating this set is part of the 
negative-weight percolation \cite{melchert2008} problem, which
 yields a globally optimal solution. This works by transforming
the problem into a \emph{minimum-weight perfect matching problem} which
is a standard problem in graph theory and 
can be solved exactly in a time growing only polynomially with system size.
Note that while the total 
edge weight of the whole set of loops is optimized, each 
individual loop has a total negative or zero weight.  
These loops separate clusters of 
aligned spins that form a GS of the system. 
Suppose two spins 
share a bond that favors antiparallel alignment, then the 
corresponding edge in $G'$ that separates these spins carries a 
negative weight. Because we look for minimal weighted loops, this 
edge is likely to be included in one of the loops and thus the 
spins are in different clusters, fulfilling the antiparallel 
alignment of the bond. 
In general, the weight of a loop
is the negative of the 
energy of the DW surrounding the spins inside the loop.
Therefore, by flipping the spins inside a loop, the total energy will be 
decreased by twice this amount.

Therefore, the
last step is to construct a ground state from the minimal weighed loops. This is achieved by arbitrarily setting the top left spin to \textit{up}. The orientation of neighboring spins can then be determined by looking at the loops. If spins are separated by a loop they need be aligned in opposite direction. This process is repeated until all spin have been assigned an orientation. Since there is no external field, the configuration obtained by a global spin flip is also a GS of the system.

From the calculated GS, we can easily obtain the
magnetization values of the sublattices. Nevertheless, 
due to the discrete structure of the model, the GS is highly degenerate.
Therefore, we used a small randomization of the bonds to lift the
degeneracy. Instead of the original bonds values $J_{i,j}=0,-1$, we
used 
\begin{equation}
\hat{J}_{i, j} = S \, J_{i, j} + X_{i,j}\,,\label{eq:randomization}
\end{equation}
 with a constant scaling 
factor $S$ and uniform distributed discrete random variables 
$X_{i,j} \in \{-V,-V+1,\ldots, V-1, V\} \subset \mathbb{Z}$. The values 
$S = 10^6$  and $V = 100$ are used throughout the remaining analysis. The large 
scaling factor is used to keep the bond strength an integer, while 
allowing for slight variations such that the ground state of the
modified system is also a GS of the original system, for each realization.
 Although this randomization does not
guarantee a uniform sampling of the ground states, it was shown
\cite{amoruso2004} that for the two-dimensional random-bond Ising model,
the influence of the bias is very weak such that the results are
reliable within the statistical error bars.

Furthermore, we also studied the scaling of domain-wall energies. 
The DWs are induced 
\cite{mcmillan1984,bray1984,saul1993,rieger1996,hartmann2001} 
for a given realization
by first calculating the GS for the original system, leading
to a ground state energy $E_p$. Another GS is obtained for
a modified system such that a domain wall is induced. Here, the second GS
 with energy $E_{ap}$ 
is calculated for a system,
where the boundary conditions are switched from periodic to antiperiodic
in the horizontal direction. The switch of the boundary conditions is
realized by inverting the sign of the bonds in one (top-bottom) column of
bonds. The changed boundary conditions induce that in the GS the spins left and
right of the switched bonds obtain relative orientations 
opposite to the
ground states. Due to the periodicity in the horizontal direction,
there must be another line where the relative orientation of the spins across 
the line is opposite to the relative orientation in the GS of the 
original system. This is the domain
wall generated by this procedure. The energy of the domain wall is given by
\begin{equation}
\Delta E =E_{\rm p}-E_{\rm ap}\,.
\end{equation}
If the disorder-averaged value $\langle \Delta E \rangle$ increases with system
size, domain walls become more and more expensive, thus an ordered
state with nonzero order parameter $m_\alpha$ is stable. 
On the other hand, if $\langle \Delta E \rangle$
decreases with system size, for large systems 
arbitrary small thermal fluctuations
will be sufficient to destroy an ordered state, which means $T_\text c=0$.
Nevertheless, a non-magnetized state might exhibit spin-glass order.
This is signified by a growth of the width $\sigma(\Delta E)$
of the (disorder) distribution of domain-wall energies  when
increasing the system size $L$, while the average
$\langle \Delta E \rangle$ decreases.  Equivalently to the width, 
one could monitor the
size dependence of the average $\langle |\Delta E| \rangle$ of the
absolute value of the domain-wall energy. Below, we will use this approach
to show that the KS model exhibits indeed a global antiferromagnetic order,
i.e., a ferromagnetic order for the fully occupied sublattices,
 for certain
values of $p$, but no spin glass order.

\section{Results}
To obtain the following results, between 5000 and 10000 
random realizations of the
disorder for the Kaya-Berker model were studied for various system sizes, 
for many values of the disorder parameter $p$. 
We studied square systems $L = L_x = L_y$ with $L\in [30,345]$.
For each realization we
calculated exact ground states of the system with periodic and 
anti-periodic boundary conditions, respectively. Each realization and 
both GSs are saved to disk and are analyzed later. All systems have open boundary conditions in the $y$-direction, since the GS algorithm cannot handle periodicity in both directions. This does not impact the ordering of the system, as the change of boundary conditions, to induce a domain wall,
 is performed perpendicular to the open boundary condition. 
The DW therefore spans the system between the open boundaries. 

\begin{figure}[htb]
 \includegraphics{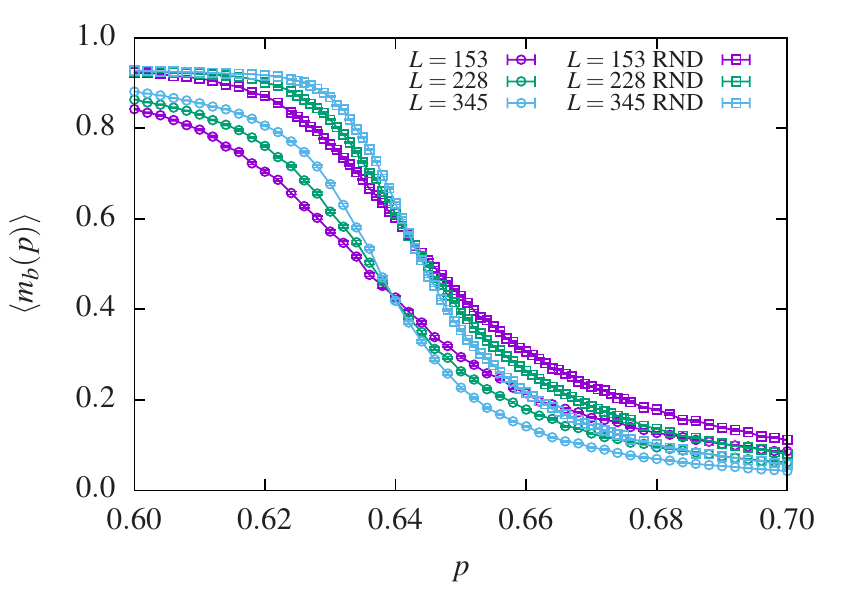}
 \caption{(Color online) Magnetization of sublattice $b$ at different fractions $p$ of occupied sites and system sizes $L$. The RND data sets include slight variations of the bond strengths according to Eq.~(\ref{eq:randomization}).
 \label{fig:mag_fail}}
\end{figure}

First the magnetization of the Kaya-Berker model in the ground state is studied. The magnetization \eqref{eq:magnetization} is calculated per 
lattice and plotted as a function of the fraction of 
occupied spins $p$.  The sign of the sublattice magnetization is 
chosen so that magnetization for lattice $b$ is always positive. One problem with calculating GSs is that most observables depend on the specific GS that is generated. The discrete nature of the Kaya-Berker model results in exponentially many GSs, all sharing the same energy, but varying in other properties. This could lead to biased results if the GS algorithm favors specific states. In fact this is the case for the presented GS algorithm, as visible in 
Fig.~\ref{fig:mag_fail}, where we compare the resulting magnetizations
for the original bonds and for bonds slightly randomized according
to Eq.~(\ref{eq:randomization}).
All the magnetization values without bond randomization are clearly 
lower than the ones with randomization. The intersection of the 
curves also shifts slightly to the right when bond randomization is used. 
As mentioned above, our final results are obtained for the
slightly randomized realizations.

Next, the critical point is determined by performing a finite-size
scaling  analysis of the 
Binder parameter \cite{binder1981} 
\begin{equation}
b_\alpha = \frac 1 2 \left(
3- \frac{\langle m_\alpha^4 \rangle}{\langle m_\alpha^2 \rangle^2}
\right)
\end{equation}
for the data of the slightly randomized
systems. When plotting $b$ as a function of the disorder parameter $p$,
the curves for different system sizes $L$ will intersect (for large
enough system sizes) at the critical point $p_c$ where
the sub lattice ferromagnetic (global antiferromagnetic) 
order disappears. This allows
for a convenient determination of the critical point. Furthermore, finite-size
scaling \cite{cardy1988} shows that when rescaling the $p$ axis according 
to $(p-p_c)L^{1/\nu}$, for the correct value of $p_c$ and
a suitably chosen value of $\nu$,
 the data will collapse onto a single curve. 
More precisely, the Binder parameter follows
\begin{equation}
b(p,L)=\tilde b\left((p-p_c)L^{1/\nu}\right)\,,
\end{equation} 
where $\tilde b(\ldots)$ is a non-size-dependent
function of one scaled variable. The quantity $\nu$
is a \emph{critical exponent} which describes the divergence of the correlation
length when approaching a second-order phase transition. The actual value
of $\nu$ (together with other critical exponents) allows
 to classify second-order phase transitions according to 
\emph{universality classes}.
\begin{figure}
 \centering
  \includegraphics[width=0.9\columnwidth]{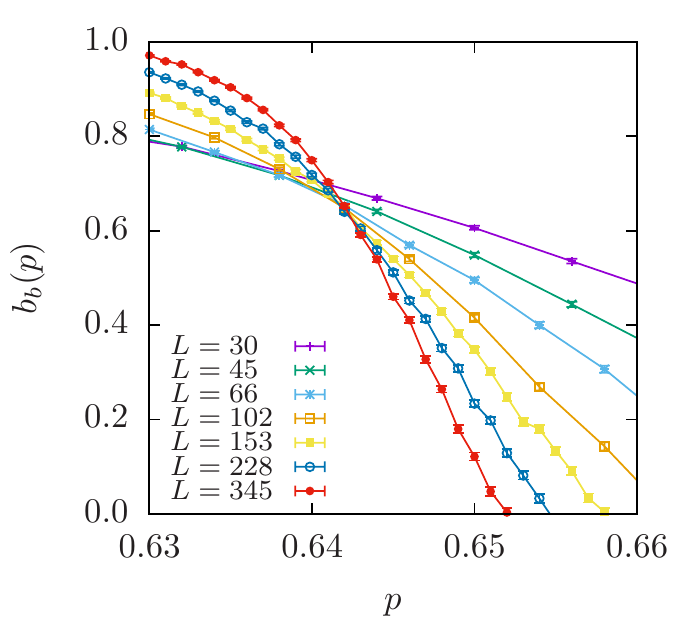}
 \caption{(Color online) Binder parameter of sublattice $b$ as 
a function of the sublattice occupancy $p$, for
different system sizes. 
 \label{fig:binder}}
\end{figure}
\begin{figure}
 \centering
  \includegraphics[width=0.9\columnwidth]{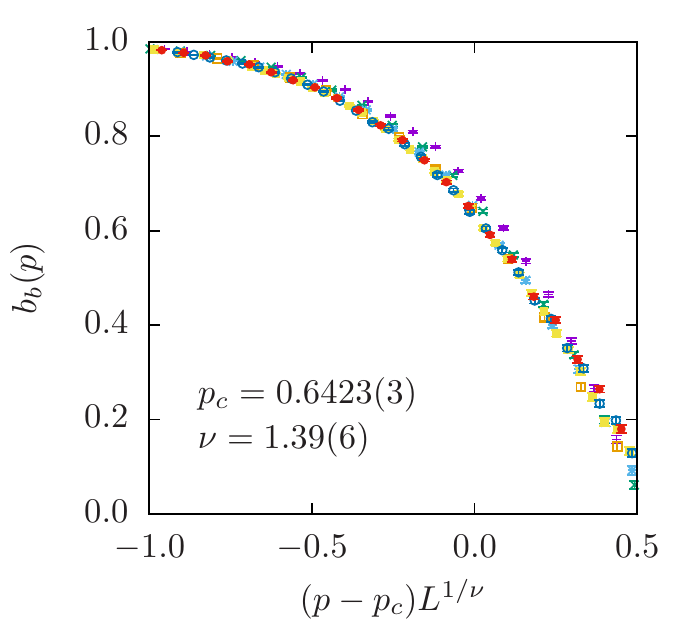}
 \caption{(Color online) Rescaled Binder parameter with a rescaled $p$-axis
using the appropriate scaling variables $p_c=0.6423(3)$, $\nu=1.39$. 
 \label{fig:binder:scaled}}
\end{figure}

The results for the Binder parameter and the best data
collapse, as obtained from the \emph{autoscale script}
\cite{autoscale} are shown in Figs.~\ref{fig:binder} and 
\ref{fig:binder:scaled}. The $x$-range
of the collapse was restricted to $[-0.75, 0.25]$, which gave the best
collapse quality of $S = 1.17$. $S$ describes the mean-squared
fluctuation of the data along the collapse curve measured in
\emph{autoscale} in terms of error bars. The system sizes $L = 30$
and $L = 45$ were excluded from the data collapse, since they deviate
from the other curves and resulted in a worse collapse. This is
because of their quite small system size, which would require
additional corrections to match the other curves. We find that the
Kaya-Berker model has a critical point of $p_\text c = 0.6423(3)$.
This result is much smaller than any of the previous results of
$p_\text c = 0.958$ by Kaya and Berker \cite{kaya2000}, $p_\text c
\approx 0.95$ by using a Monte Carlo simulation \cite{robinson2003},
or $p_\text c = 0.875$ obtained using effective-field theory (EFT)
\cite{zukovic2012}. However, the discrepancy can be explained by
taking a closer look at the previously used methods. 
First, the results by Kaya and
Berker are obtained by HSMF theory. This method
includes several approximations, like the mean-field nature of the
approach and the partial use of locally averaged magnetizations. Furthermore,
 this method includes a form
of stochastic iterations. These iterations often gets stuck in local
minima. In fact, the authors found a multiplicity of solutions and used only
the most stable set to determine the critical point. Other not so
stable solutions are fragmented and show a much lower magnetization
than the stable one. Indeed the solution with the lowest magnetization
looks like it could become zero somewhere around $0.6 < p < 0.7$,
which would coincide with our $p_\text c = 0.6423(3)$ result. The
other result obtained by MC simulations also suffers from the problem
that the dynamic of the Kaya-Berker model becomes very slow at low
temperatures and often also gets stuck in local energy minima. Last, the
other approaches, which are very similar to the 
further approximation of
HSMF, impose also sublattice-wise uniformity of the magnetization. This
restriction may not capture the whole behavior of the model, leading
to wrong results. Therefore, the exact GS approach which we used here
is much more
reliable than any of the previous results since it neither includes
approximations nor it does suffer from convergence problems. Finally,
we can treat much larger sizes compared to the previous approaches.

Although we are here not mainly interested in characterizing the
phase transition, we also obtained an estimate $\nu=1.39(6)$ for the
value of the critical exponent of the correlation length. No corresponding
result for the Kaya-Berker model 
for this critical exponent at the antiferromagnet-paramagnet
transition is known to us in the literature. Thus a direct comparison
is not possible. Nevertheless, for the random-bond Ising model, which exhibits
a ferromagnet-paramagnet transition 
(with spin-glass behavior at exactly $T=0$), a value $\nu=1.55(1)$ 
was found \cite{amoruso2004}. 
This is not fully compatible, but only two error bars away
from the value obtained here. Therefore, the transitions might be
in the same universality class. Anyway, we proceed towards our main
aim of the paper, to show that no thermodynamic stable spin-glass
phase exists.

\begin{figure}[htb]
 \centering
 \begin{minipage}[t]{0.3\columnwidth}
  \includegraphics[width=1\textwidth]{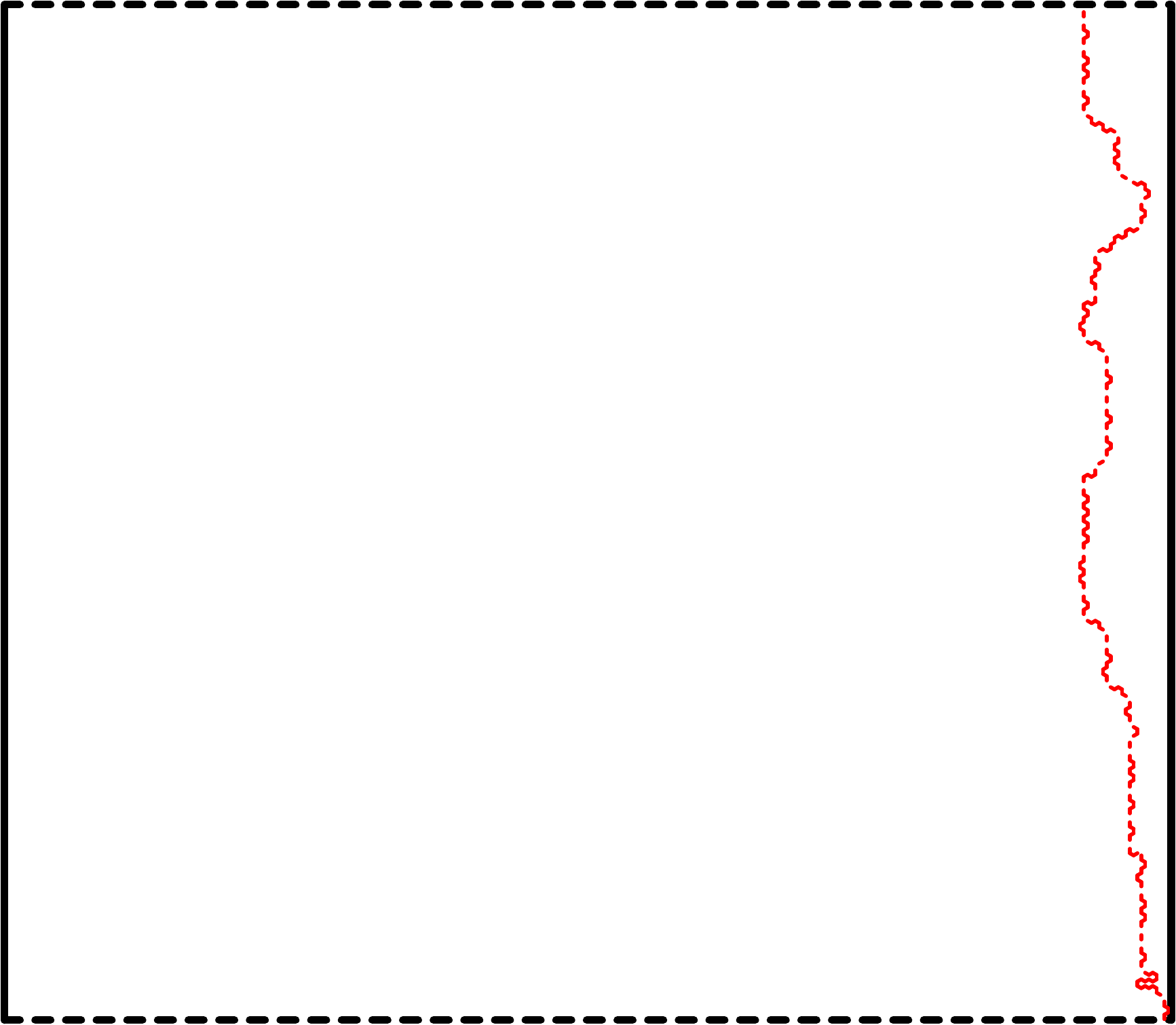}
 \end{minipage}
 \hspace{0.2cm}
 \begin{minipage}[t]{0.3\columnwidth}
  \includegraphics[width=1\textwidth]{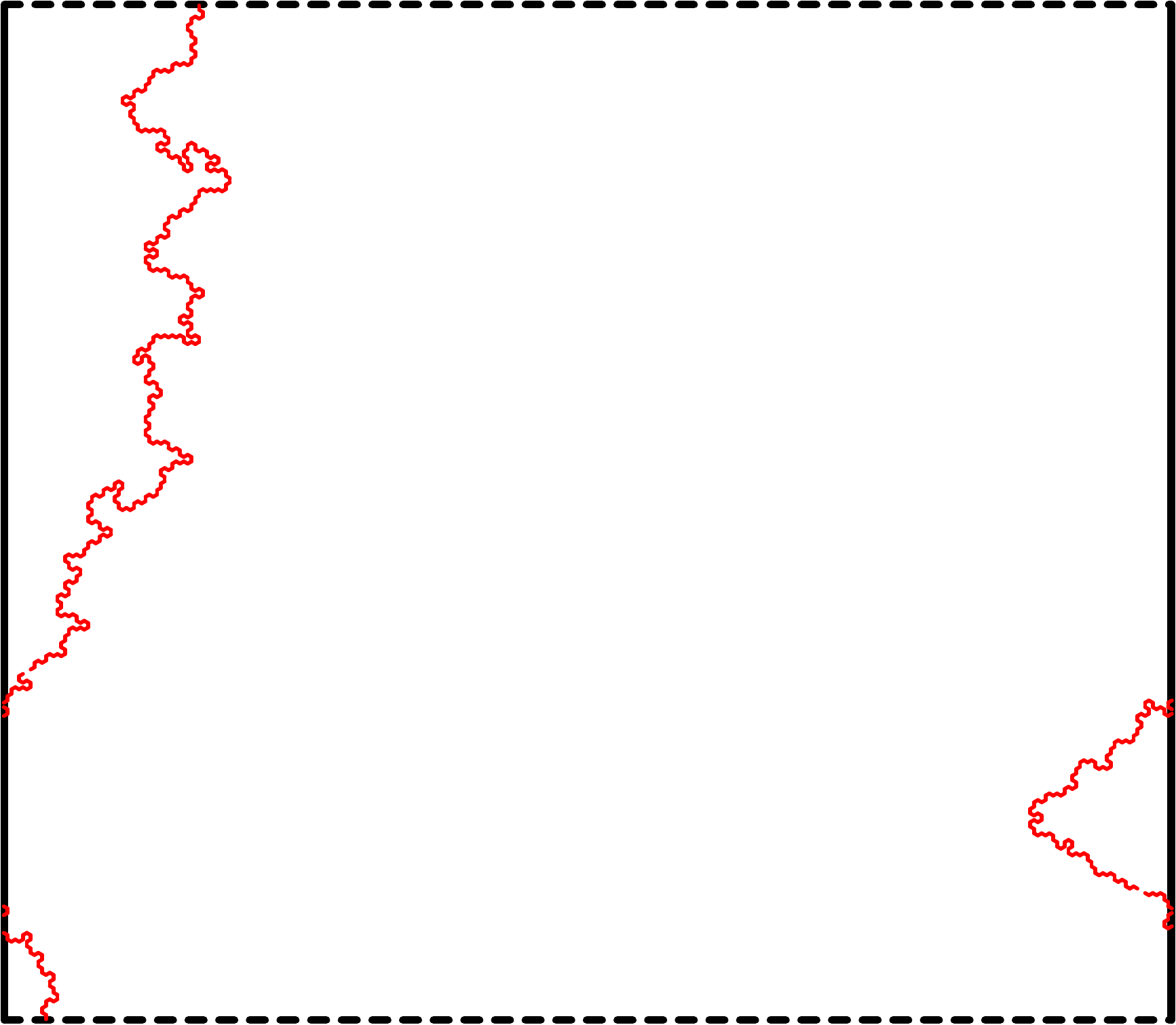}
 \end{minipage}
 \hspace{0.2cm}
 \begin{minipage}[t]{0.3\columnwidth}
  \includegraphics[width=1\textwidth]{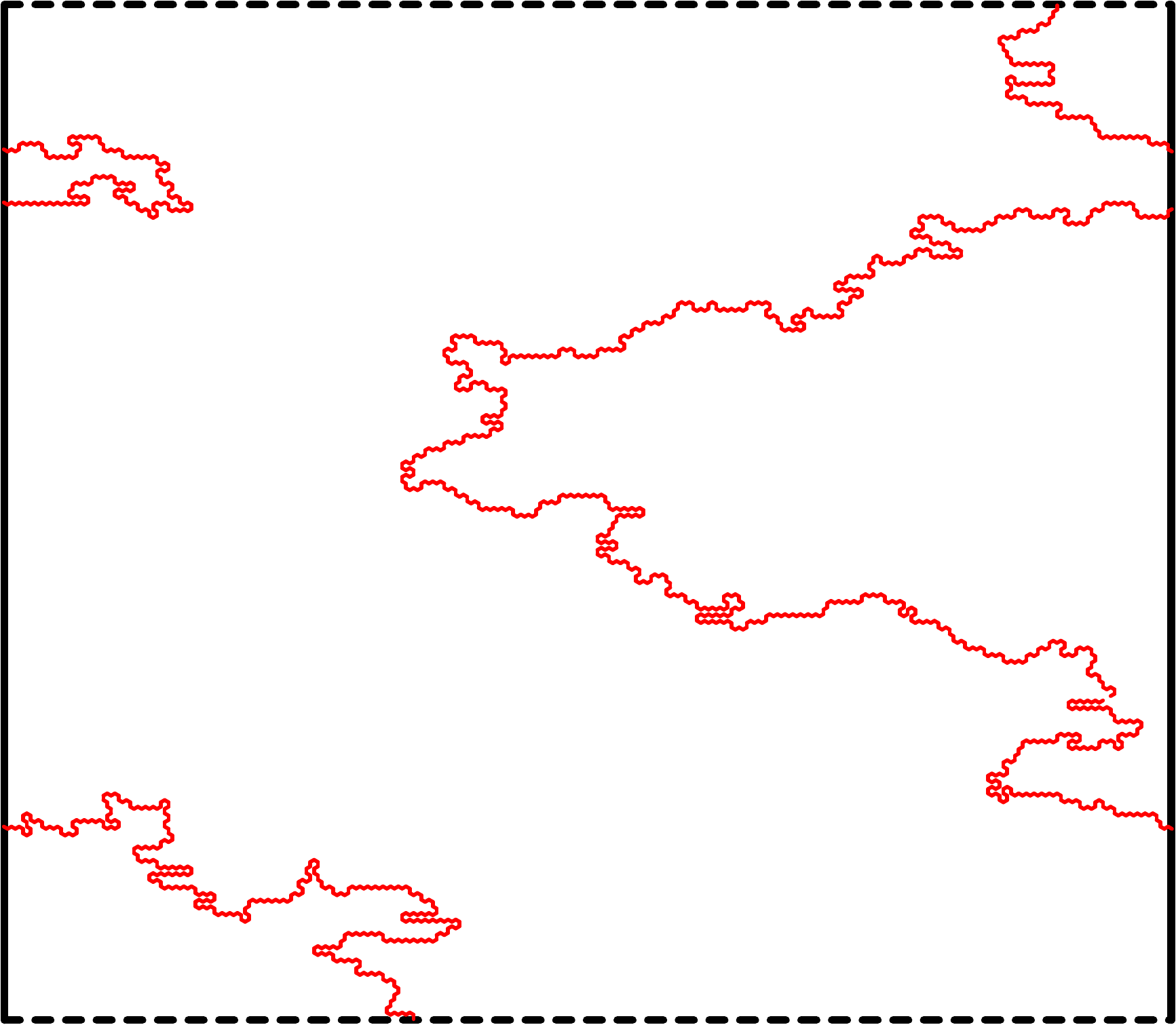}
 \end{minipage}
 \caption{(Color online) Exemplary domain walls at different 
occupancy. The solid and dashed border marks periodic and open 
boundary conditions respectively. The occupancy was set to 
$p=0.2$ (left), $p=0.6$ (center) and $p=0.8$ (right).}
 \label{fig:domain_walls}
\end{figure}

Next, the domain wall length $l$ is studied. This is another
property which strongly depends on the specific GS of a given
realization. As explained above, DWs
 are obtained by calculating the GS
of a  realization, flipping the boundary condition,
calculating a new GS, and comparing the two obtained GS.
Note that
 the bond randomization is again used, which should result in typical domain
wall lengths among many possible degenerate DWs. This is sufficient
for the present purpose of identifying the transition point
where the order disappears.
 A more sophisticated method to determine shortest and
maximal length domain walls is presented in Ref.~\cite{melchert2007}. This
analysis uses $10,000$ samples for each system size $L\in [30,777]$ 
and occupancy
value $p$. Some exemplary domain walls at different occupancy $p$ are
shown in Fig.~\ref{fig:domain_walls}. For small values of $p$,
where the GS is ordered, the DWs are very straight. With increasing 
value of $p$ the domain walls become more fractal.

\begin{figure}
 \centering
  \includegraphics[width=\columnwidth]{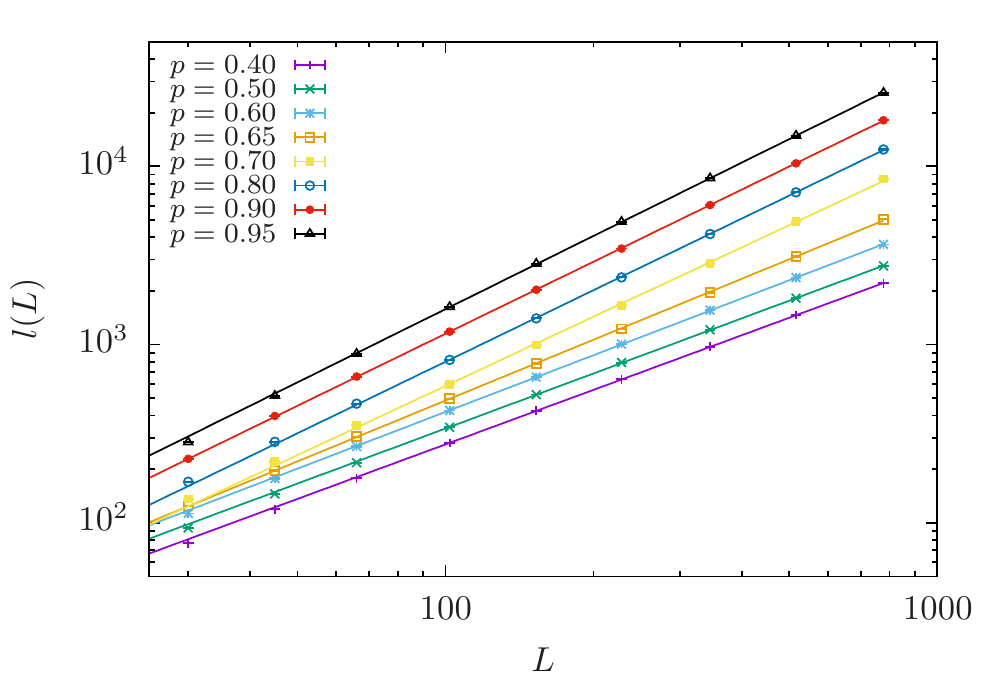}
 \caption{(Color online) Domain wall length $l$ at different system
sizes $L$. The straight lines are power-law fits to the
data sets.  }
 \label{fig:length}
\end{figure}

\begin{figure}
  \includegraphics[width=\columnwidth]{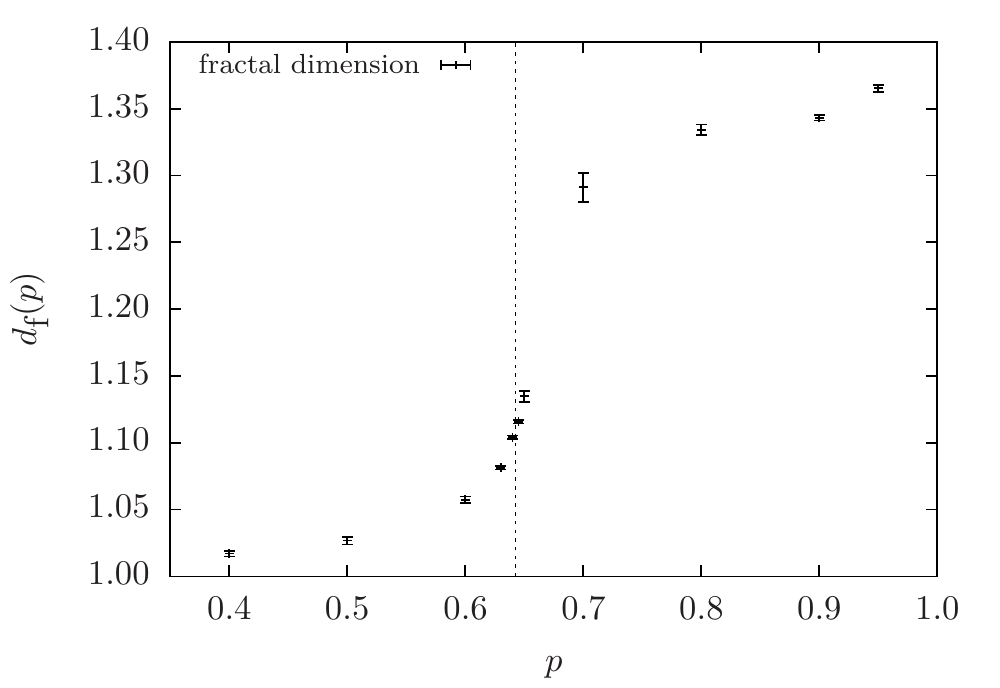}
 \caption{(Color online) Fractal dimension $d_\text f$ of the domain wall
length at different occupancy $p$. The vertical dashed line marks the
critical point $p_\text c = 0.6423(3)$. }
 \label{fig:length:exp}
\end{figure}

The measured averaged lengths $l$ 
of the DWs are shown in Fig.~\ref{fig:length} as a function
of the system size $L$. All
of the data sets show a clean power-law behavior of the form
\begin{equation}
l(L) \sim L^{d_{\rm f}}\,.
\end{equation}
where $d_f$ is the fractal exponent which depends on $p$.
A power-law fit was
performed for all data sets, excluding the small system sizes $L <
50$. The fits match the data sets very well. The resulting
fractal dimension $d_\text f$,
see Fig.~\ref{fig:length:exp}, exhibits a change between the $p \le
0.65$ and $p \ge 0.70$ data sets visible as a rather sharp jump in the
plot. Apart from this sharp jump, the fractal dimension also grows
slowly with the occupancy. By interpolating between the two values
closest to the critical point at $p = 0.64$ and $p = 0.645$, the
fractal dimension at the critical point is $d_\text f =
1.109(2)$. This value is different from the value $d_\text f =
1.222(1)$ obtained for the 2D random-bond Ising model on a triangular lattice
\cite{melchert2011} and also on a square lattice \cite{melchert2009}. 

\begin{figure}[htb]
 \centering
 \includegraphics{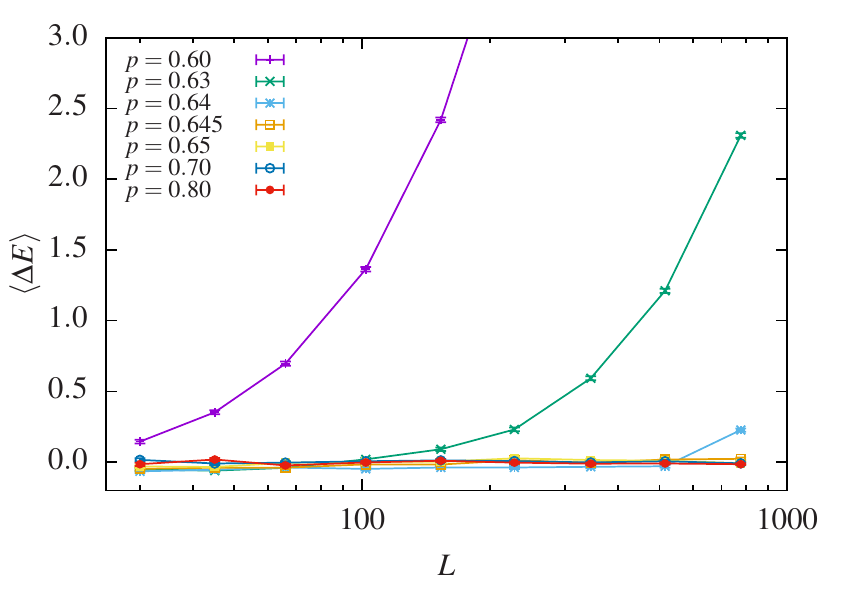}
 \caption{(Color online) Average domain wall energy at different 
system sizes. Data points are connected using straight lines for 
better visibility.}
 \label{fig:delta}
\end{figure}

The property which is discussed last is the DW energy. The same 
$10,000$ samples used for the domain wall length are analyzed 
here. This results in the data shown in Fig.~\ref{fig:delta}.
 The data was plotted on a semi-log scale, since some
average values are negative. For occupancy $p < 0.645$ the average DW
energy increases linearly. This includes values of $p < 0.6$ which
were omitted from the plot for a clearer image. This indicates a
sub lattice ferromagnetic, i.e., globally antiferromagnetic, phase, 
where introduction of a DW 
breaks the long range order and costs energy. 
This
behavior is expected for occupancy smaller than the critical point
$p_\text c$, since the system is antiferromagnetic at $p = 0$. 
For $p \ge
0.645$ the DW energy is roughly zero, which is typical for 
both possibilities of a paramagnetic and a spin-glass
phase. Note that fairly large system sizes are required, as the $p =
0.63$ curve is also approximately zero at first, but then increases
around $L = 100$. Similarly, the $p = 0.64$ curve only deviates from
zero around $L = 600$. This means that systems around these occupancy
values may appear not sub lattice ferromagnetically ordered
at small system sizes. Further
transitions at greater values of
$p$ and much larger systems sizes cannot be
fully excluded from this plot, but are unlikely, as we found the
critical point to be $p_\text c = 0.6423(3)$.

\begin{figure}[htb]
 \centering
 \includegraphics{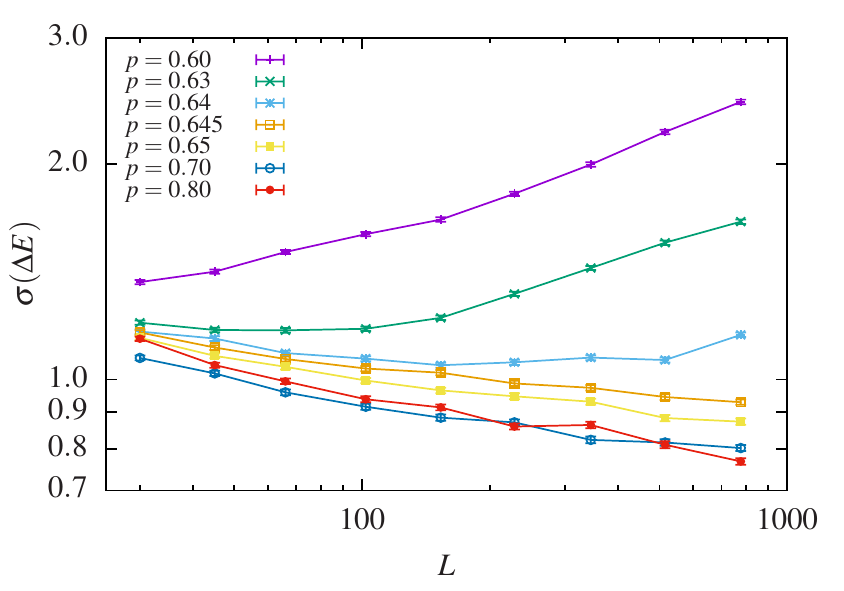}
 \caption{(Color online) Standard deviation of domain wall energy 
at different system sizes. Data points are connected using straight 
lines for better visibility.}
 \label{fig:delta_std}
\end{figure}

To determine whether there exists a spin-glass phase,
 we look at the standard 
deviation of the DW energy in Fig.~\ref{fig:delta_std}.
 Again, the curves split into two categories: Increasing
standard deviation with the system size at $p < 0.645$ and decreasing
values at $p \ge 0.645$.  This suggests that the ferromagnetic phase
is stable at finite temperatures, but the claimed spin-glass phase
only exists at $T = 0$. For a stable spin-glass phase the standard
deviation would have
 to increase in some region of
the $p \ge 0.645$ range, which is clearly
not observed. With these results it can be concluded that previous
reports of a stable $T_\text c > 0$ spin-glass phase are not
supported by our results.

\rule{0pt}{1pt}

\section{Conclusions}

We have numerically studied  
the Kaya-Berker model, with a site occupancy $p$ of one 
sublattice,  by using an exact ground state
algorithm. Thus, we were able to obtain $T=0$ results in exact equilibrium.
Since the ground-state calculation is equivalent to obtain a minimum-weight
perfect matching, it is possible to obtain ground states with a running
time growing only polynomially in system size. Therefore, we were able
to study very large system sizes of up to $\sim 10^5$ lattice sites.

From the obtained GSs, we calculated the magnetization and the corresponding
Binder parameter as a function of the occupancy $p$.
 Here we used a slight randomization of the bonds
to obtain a almost unbiased sampling of the degenerate ground states.
From the Binder parameter, we obtained a critical value $p_c=0.6423(3)$ where
the sublattice magnetization vanishes. This value is considerably smaller
than previous results. Nevertheless, the previous results were obtained
by using methods which involve approximations, non-equilibrium
sampling often without a guarantee of convergence and they were
restricted to small system sizes. Therefore, our results are much more
reliable than those obtained from the previous studies. Although there 
might be a slight bias from the sampling of the ground states,
from the comparison of the results for the pure samples and the randomized
samples, as well as from the previous results for the random bond (spin-glass)
model, we know that his influence is very small, much smaller than
the discrepancy to the previous results.

We also studied domain walls which are
obtained by comparing the GS of the original realizations with
periodic boundary conditions in $x$-direction and the GS of the
realization with antiperiodic boundary conditions. By analyzing the length
of the domain-walls and its dependence on the system size, we obtained
the fractal dimension. The result for
the fractal dimension as a function of the occupancy shows a fractal
dimension of basically one in the phase where the sublattice magnetization
is non-zero. Close to the  point a strong increase of the
fractal dimension can be observed, supporting the results for the
phase transition obtained from the
magnetization.

The results for scaling of the
average DW energy are also compatible with a transition at 
the location $p_c$ where the
magnetization vanishes. The variance of the DW energy
distribution changes from growing to shrinking  with
the system size $L$ at (or close to) the same point $p_c$.
Thus, it is rather likely, that the Kaya-Berker model does not
exhibit a spin-glass phase at a low but finite temperature, because this
would have to go together with a decrease of the mean value and an increase
of the variance with the system size at the same time at some values of $p$.

To conclude, so far for no two-dimensional
(random) frustrated Ising systems with
short-range nearest neighbor interactions a low-temperature
spin-glass phase with a finite critical
temperature has been found (although for some it had been claimed before).
To conclude, it still remains an open question whether such a system
exists, but is appears to be unlikely to the authors.

\begin{acknowledgments}
The simulations were performed on the HPC clusters HERO
and CARL  of the University
of Oldenburg jointly funded by the DFG 
through its Major Research Instrumentation Programme (INST 184/108-1 FUGG
and INST 184/157-1 FUGG) and the
Ministry of Science and Culture (MWK) of the Lower Saxony State.
\end{acknowledgments}

\bibliography{references}

\begin{thebibliography}{36}
\expandafter\ifx\csname natexlab\endcsname\relax\def\natexlab#1{#1}\fi
\expandafter\ifx\csname bibnamefont\endcsname\relax
  \def\bibnamefont#1{#1}\fi
\expandafter\ifx\csname bibfnamefont\endcsname\relax
  \def\bibfnamefont#1{#1}\fi
\expandafter\ifx\csname citenamefont\endcsname\relax
  \def\citenamefont#1{#1}\fi
\expandafter\ifx\csname url\endcsname\relax
  \def\url#1{\texttt{#1}}\fi
\expandafter\ifx\csname urlprefix\endcsname\relax\def\urlprefix{URL }\fi
\providecommand{\bibinfo}[2]{#2}
\providecommand{\eprint}[2][]{\url{#2}}

\bibitem[{\citenamefont{Young}(1997)}]{young1998}
\bibinfo{editor}{\bibfnamefont{A.~P.} \bibnamefont{Young}}, ed.,
  \emph{\bibinfo{title}{Spin Glasses and Random Fields}}
  (\bibinfo{publisher}{World Scientific}, \bibinfo{address}{Singapore},
  \bibinfo{year}{1997}).

\bibitem[{\citenamefont{Hartmann}(2015)}]{practical_guide2015}
\bibinfo{author}{\bibfnamefont{A.~K.} \bibnamefont{Hartmann}},
  \emph{\bibinfo{title}{{Big Practical Guide to Computer Simulations}}}
  (\bibinfo{publisher}{World Scientific}, \bibinfo{address}{Singapore},
  \bibinfo{year}{2015}).

\bibitem[{\citenamefont{Newman and Barkema}(1999)}]{newman1999}
\bibinfo{author}{\bibfnamefont{M.~E.~J.} \bibnamefont{Newman}}
  \bibnamefont{and} \bibinfo{author}{\bibfnamefont{G.~T.}
  \bibnamefont{Barkema}}, \emph{\bibinfo{title}{Monte Carlo Methods in
  Statistical Physics}} (\bibinfo{publisher}{Oxford University Press},
  \bibinfo{year}{1999}).

\bibitem[{\citenamefont{Hartmann and Rieger}(2001)}]{opt-phys2001}
\bibinfo{author}{\bibfnamefont{A.~K.} \bibnamefont{Hartmann}} \bibnamefont{and}
  \bibinfo{author}{\bibfnamefont{H.}~\bibnamefont{Rieger}},
  \emph{\bibinfo{title}{Optimization Algorithms in Physics}}
  (\bibinfo{publisher}{Wiley-VCH}, \bibinfo{address}{Weinheim},
  \bibinfo{year}{2001}).

\bibitem[{\citenamefont{Hartmann and Rieger}(2004)}]{opt-phys2004}
\bibinfo{editor}{\bibfnamefont{A.~K.} \bibnamefont{Hartmann}} \bibnamefont{and}
  \bibinfo{editor}{\bibfnamefont{H.}~\bibnamefont{Rieger}}, eds.,
  \emph{\bibinfo{title}{New Optimization Algorithms in Physics}}
  (\bibinfo{publisher}{Wiley-VCH}, \bibinfo{address}{Weinheim},
  \bibinfo{year}{2004}).

\bibitem[{\citenamefont{McMillan}(1984)}]{mcmillan1984}
\bibinfo{author}{\bibfnamefont{W.~L.} \bibnamefont{McMillan}},
  \bibinfo{journal}{Phys. Rev. B} \textbf{\bibinfo{volume}{30}},
  \bibinfo{pages}{476} (\bibinfo{year}{1984}).

\bibitem[{\citenamefont{Bray and Moore}(1984)}]{bray1984}
\bibinfo{author}{\bibfnamefont{A.~J.} \bibnamefont{Bray}} \bibnamefont{and}
  \bibinfo{author}{\bibfnamefont{M.~A.} \bibnamefont{Moore}},
  \bibinfo{journal}{J. Phys. C} \textbf{\bibinfo{volume}{17}},
  \bibinfo{pages}{L463} (\bibinfo{year}{1984}).

\bibitem[{\citenamefont{Saul and Kardar}(1993)}]{saul1993}
\bibinfo{author}{\bibfnamefont{L.}~\bibnamefont{Saul}} \bibnamefont{and}
  \bibinfo{author}{\bibfnamefont{M.}~\bibnamefont{Kardar}},
  \bibinfo{journal}{Phys. Rev. E} \textbf{\bibinfo{volume}{48}},
  \bibinfo{pages}{R3221} (\bibinfo{year}{1993}).

\bibitem[{\citenamefont{Rieger et~al.}(1996)\citenamefont{Rieger, Santen,
  Blasum, Diehl, J\"unger, and Rinaldi}}]{rieger1996}
\bibinfo{author}{\bibfnamefont{H.}~\bibnamefont{Rieger}},
  \bibinfo{author}{\bibfnamefont{L.}~\bibnamefont{Santen}},
  \bibinfo{author}{\bibfnamefont{U.}~\bibnamefont{Blasum}},
  \bibinfo{author}{\bibfnamefont{M.}~\bibnamefont{Diehl}},
  \bibinfo{author}{\bibfnamefont{M.}~\bibnamefont{J\"unger}}, \bibnamefont{and}
  \bibinfo{author}{\bibfnamefont{G.}~\bibnamefont{Rinaldi}},
  \bibinfo{journal}{J. Phys. A} \textbf{\bibinfo{volume}{29}},
  \bibinfo{pages}{3939} (\bibinfo{year}{1996}).

\bibitem[{\citenamefont{Houdayer}(2001)}]{houdayer2001}
\bibinfo{author}{\bibfnamefont{J.}~\bibnamefont{Houdayer}},
  \bibinfo{journal}{Eur. Phys. J. B} \textbf{\bibinfo{volume}{22}},
  \bibinfo{pages}{479} (\bibinfo{year}{2001}), ISSN \bibinfo{issn}{1434-6036}.

\bibitem[{\citenamefont{Hartmann and Young}(2001)}]{hartmann2001}
\bibinfo{author}{\bibfnamefont{A.~K.} \bibnamefont{Hartmann}} \bibnamefont{and}
  \bibinfo{author}{\bibfnamefont{A.~P.} \bibnamefont{Young}},
  \bibinfo{journal}{Phys. Rev. B} \textbf{\bibinfo{volume}{64}},
  \bibinfo{pages}{180404} (\bibinfo{year}{2001}).

\bibitem[{\citenamefont{Carter et~al.}(2002)\citenamefont{Carter, Bray, and
  Moore}}]{carter2002}
\bibinfo{author}{\bibfnamefont{A.~C.} \bibnamefont{Carter}},
  \bibinfo{author}{\bibfnamefont{A.~J.} \bibnamefont{Bray}}, \bibnamefont{and}
  \bibinfo{author}{\bibfnamefont{M.~A.} \bibnamefont{Moore}},
  \bibinfo{journal}{Phys. Rev. Lett.} \textbf{\bibinfo{volume}{88}},
  \bibinfo{pages}{077201} (\bibinfo{year}{2002}).

\bibitem[{\citenamefont{Hartmann}(2003)}]{hartmann2003}
\bibinfo{author}{\bibfnamefont{A.~K.} \bibnamefont{Hartmann}},
  \bibinfo{journal}{Phys. Rev. B} \textbf{\bibinfo{volume}{67}},
  \bibinfo{pages}{214404} (\bibinfo{year}{2003}).

\bibitem[{\citenamefont{Kaya and Berker}(2000)}]{kaya2000}
\bibinfo{author}{\bibfnamefont{H.}~\bibnamefont{Kaya}} \bibnamefont{and}
  \bibinfo{author}{\bibfnamefont{A.~N.} \bibnamefont{Berker}},
  \bibinfo{journal}{Phys. Rev. E} \textbf{\bibinfo{volume}{62}},
  \bibinfo{pages}{R1469} (\bibinfo{year}{2000}).

\bibitem[{\citenamefont{Wannier}(1950)}]{wannier1950}
\bibinfo{author}{\bibfnamefont{G.~H.} \bibnamefont{Wannier}},
  \bibinfo{journal}{Phys. Rev.} \textbf{\bibinfo{volume}{79}},
  \bibinfo{pages}{357} (\bibinfo{year}{1950}).

\bibitem[{\citenamefont{Houtappel}(1950)}]{houtappel1950}
\bibinfo{author}{\bibfnamefont{R.~M.~F.} \bibnamefont{Houtappel}},
  \bibinfo{journal}{Physica} \textbf{\bibinfo{volume}{16}}, \bibinfo{pages}{425
  } (\bibinfo{year}{1950}), ISSN \bibinfo{issn}{0031-8914}.

\bibitem[{\citenamefont{Grest and Gabl}(1979)}]{grest1979}
\bibinfo{author}{\bibfnamefont{G.~S.} \bibnamefont{Grest}} \bibnamefont{and}
  \bibinfo{author}{\bibfnamefont{E.~G.} \bibnamefont{Gabl}},
  \bibinfo{journal}{Phys. Rev. Lett.} \textbf{\bibinfo{volume}{43}},
  \bibinfo{pages}{1182} (\bibinfo{year}{1979}).

\bibitem[{\citenamefont{Blackman et~al.}(1981)\citenamefont{Blackman, Kemeny,
  and Straley}}]{blackman1981}
\bibinfo{author}{\bibfnamefont{J.~A.} \bibnamefont{Blackman}},
  \bibinfo{author}{\bibfnamefont{G.}~\bibnamefont{Kemeny}}, \bibnamefont{and}
  \bibinfo{author}{\bibfnamefont{J.~P.} \bibnamefont{Straley}},
  \bibinfo{journal}{J. Phys. C} \textbf{\bibinfo{volume}{14}},
  \bibinfo{pages}{385} (\bibinfo{year}{1981}).

\bibitem[{\citenamefont{And\'erico et~al.}(1982)\citenamefont{And\'erico,
  Fern\'andez, and Streit}}]{anderico1982}
\bibinfo{author}{\bibfnamefont{C.~Z.} \bibnamefont{And\'erico}},
  \bibinfo{author}{\bibfnamefont{J.~F.} \bibnamefont{Fern\'andez}},
  \bibnamefont{and} \bibinfo{author}{\bibfnamefont{T.~S.~J.}
  \bibnamefont{Streit}}, \bibinfo{journal}{Phys. Rev. B}
  \textbf{\bibinfo{volume}{26}}, \bibinfo{pages}{3824} (\bibinfo{year}{1982}).

\bibitem[{\citenamefont{Tang et~al.}(2010)\citenamefont{Tang, Zhu, Yang, and
  Jiang}}]{tang2010}
\bibinfo{author}{\bibfnamefont{H.-L.} \bibnamefont{Tang}},
  \bibinfo{author}{\bibfnamefont{Y.}~\bibnamefont{Zhu}},
  \bibinfo{author}{\bibfnamefont{G.-H.} \bibnamefont{Yang}}, \bibnamefont{and}
  \bibinfo{author}{\bibfnamefont{Y.}~\bibnamefont{Jiang}},
  \bibinfo{journal}{Phys. Rev. E} \textbf{\bibinfo{volume}{81}},
  \bibinfo{pages}{051107} (\bibinfo{year}{2010}).

\bibitem[{\citenamefont{Robinson}(2003)}]{robinson2003}
\bibinfo{author}{\bibfnamefont{M.~D.} \bibnamefont{Robinson}}, Master's thesis,
  \bibinfo{school}{The University of Maine} (\bibinfo{year}{2003}),
  \urlprefix\url{http://digitalcommons.library.umaine.edu/etd/317}.

\bibitem[{\citenamefont{Robinson et~al.}(2011)\citenamefont{Robinson, Feldman,
  and McKay}}]{robinson2011}
\bibinfo{author}{\bibfnamefont{M.~D.} \bibnamefont{Robinson}},
  \bibinfo{author}{\bibfnamefont{D.~P.} \bibnamefont{Feldman}},
  \bibnamefont{and} \bibinfo{author}{\bibfnamefont{S.~R.} \bibnamefont{McKay}},
  \bibinfo{journal}{Chaos} \textbf{\bibinfo{volume}{21}},
  \bibinfo{pages}{037114} (\bibinfo{year}{2011}).

\bibitem[{\citenamefont{\v{Z}ukovi\v{c}
  et~al.}(2012)\citenamefont{\v{Z}ukovi\v{c}, Borovsk\'y, and
  Bob\'ak}}]{zukovic2012}
\bibinfo{author}{\bibfnamefont{M.}~\bibnamefont{\v{Z}ukovi\v{c}}},
  \bibinfo{author}{\bibfnamefont{M.}~\bibnamefont{Borovsk\'y}},
  \bibnamefont{and} \bibinfo{author}{\bibfnamefont{A.}~\bibnamefont{Bob\'ak}},
  \bibinfo{journal}{J. Magn. Magn. Mater.} \textbf{\bibinfo{volume}{324}},
  \bibinfo{pages}{2687 } (\bibinfo{year}{2012}), ISSN
  \bibinfo{issn}{0304-8853}.

\bibitem[{\citenamefont{Balcerzak et~al.}(2014)\citenamefont{Balcerzak,
  Sza\l{}owski, Ja\ifmmode \check{s}\else \v{s}\fi{}\ifmmode~\check{c}\else
  \v{c}\fi{}ur, \ifmmode \check{Z}\else \v{Z}\fi{}ukovi\ifmmode~\check{c}\else
  \v{c}\fi{}, Bob\'ak, and Borovsk\'y}}]{balcerzak2014}
\bibinfo{author}{\bibfnamefont{T.}~\bibnamefont{Balcerzak}},
  \bibinfo{author}{\bibfnamefont{K.}~\bibnamefont{Sza\l{}owski}},
  \bibinfo{author}{\bibfnamefont{M.}~\bibnamefont{Ja\ifmmode \check{s}\else
  \v{s}\fi{}\ifmmode~\check{c}\else \v{c}\fi{}ur}},
  \bibinfo{author}{\bibfnamefont{M.}~\bibnamefont{\ifmmode \check{Z}\else
  \v{Z}\fi{}ukovi\ifmmode~\check{c}\else \v{c}\fi{}}},
  \bibinfo{author}{\bibfnamefont{A.}~\bibnamefont{Bob\'ak}}, \bibnamefont{and}
  \bibinfo{author}{\bibfnamefont{M.}~\bibnamefont{Borovsk\'y}},
  \bibinfo{journal}{Phys. Rev. E} \textbf{\bibinfo{volume}{89}},
  \bibinfo{pages}{062140} (\bibinfo{year}{2014}).

\bibitem[{\citenamefont{\v{Z}ukovi\v{c}
  et~al.}(2010)\citenamefont{\v{Z}ukovi\v{c}, Borovsk\'y, and
  Bob\'ak}}]{zukovic2010}
\bibinfo{author}{\bibfnamefont{M.}~\bibnamefont{\v{Z}ukovi\v{c}}},
  \bibinfo{author}{\bibfnamefont{M.}~\bibnamefont{Borovsk\'y}},
  \bibnamefont{and} \bibinfo{author}{\bibfnamefont{A.}~\bibnamefont{Bob\'ak}},
  \bibinfo{journal}{Physics Letters A} \textbf{\bibinfo{volume}{374}},
  \bibinfo{pages}{4260 } (\bibinfo{year}{2010}), ISSN
  \bibinfo{issn}{0375-9601}.

\bibitem[{\citenamefont{Hartmann}(1999)}]{hartmann1999}
\bibinfo{author}{\bibfnamefont{A.~K.} \bibnamefont{Hartmann}},
  \bibinfo{journal}{Phys. Rev. E} \textbf{\bibinfo{volume}{59}},
  \bibinfo{pages}{84} (\bibinfo{year}{1999}).

\bibitem[{\citenamefont{Amoruso et~al.}(2003)\citenamefont{Amoruso, Marinari,
  Martin, and Pagnani}}]{amoruso2003}
\bibinfo{author}{\bibfnamefont{C.}~\bibnamefont{Amoruso}},
  \bibinfo{author}{\bibfnamefont{E.}~\bibnamefont{Marinari}},
  \bibinfo{author}{\bibfnamefont{O.~C.} \bibnamefont{Martin}},
  \bibnamefont{and} \bibinfo{author}{\bibfnamefont{A.}~\bibnamefont{Pagnani}},
  \bibinfo{journal}{Phys. Rev. Lett.} \textbf{\bibinfo{volume}{91}},
  \bibinfo{pages}{087201} (\bibinfo{year}{2003}).

\bibitem[{\citenamefont{Rom\'a et~al.}(2007)\citenamefont{Rom\'a, Risau-Gusman,
  Ramirez-Pastor, Nieto, and Vogel}}]{roma2007}
\bibinfo{author}{\bibfnamefont{F.}~\bibnamefont{Rom\'a}},
  \bibinfo{author}{\bibfnamefont{S.}~\bibnamefont{Risau-Gusman}},
  \bibinfo{author}{\bibfnamefont{A.~J.} \bibnamefont{Ramirez-Pastor}},
  \bibinfo{author}{\bibfnamefont{F.}~\bibnamefont{Nieto}}, \bibnamefont{and}
  \bibinfo{author}{\bibfnamefont{E.~E.} \bibnamefont{Vogel}},
  \bibinfo{journal}{Phys. Rev. B} \textbf{\bibinfo{volume}{75}},
  \bibinfo{pages}{020402} (\bibinfo{year}{2007}).

\bibitem[{\citenamefont{Melchert and Hartmann}(2011)}]{melchert2011}
\bibinfo{author}{\bibfnamefont{O.}~\bibnamefont{Melchert}} \bibnamefont{and}
  \bibinfo{author}{\bibfnamefont{A.~K.} \bibnamefont{Hartmann}},
  \bibinfo{journal}{Comput. Phys. Commun.} \textbf{\bibinfo{volume}{182}},
  \bibinfo{pages}{1828 } (\bibinfo{year}{2011}), ISSN
  \bibinfo{issn}{0010-4655}.

\bibitem[{\citenamefont{Melchert and Hartmann}(2008)}]{melchert2008}
\bibinfo{author}{\bibfnamefont{O.}~\bibnamefont{Melchert}} \bibnamefont{and}
  \bibinfo{author}{\bibfnamefont{A.~K.} \bibnamefont{Hartmann}},
  \bibinfo{journal}{New J. Phys.} \textbf{\bibinfo{volume}{10}},
  \bibinfo{pages}{043039} (\bibinfo{year}{2008}).

\bibitem[{\citenamefont{Amoruso and Hartmann}(2004)}]{amoruso2004}
\bibinfo{author}{\bibfnamefont{C.}~\bibnamefont{Amoruso}} \bibnamefont{and}
  \bibinfo{author}{\bibfnamefont{A.~K.} \bibnamefont{Hartmann}},
  \bibinfo{journal}{Phys. Rev. B} \textbf{\bibinfo{volume}{70}},
  \bibinfo{pages}{134425} (\bibinfo{year}{2004}).

\bibitem[{\citenamefont{Binder}(1981)}]{binder1981}
\bibinfo{author}{\bibfnamefont{K.}~\bibnamefont{Binder}}, \bibinfo{journal}{Z.
  Phys. B} \textbf{\bibinfo{volume}{43}}, \bibinfo{pages}{119}
  (\bibinfo{year}{1981}), ISSN \bibinfo{issn}{1431-584X}.

\bibitem[{\citenamefont{Cardy}(1988)}]{cardy1988}
\bibinfo{author}{\bibfnamefont{J.}~\bibnamefont{Cardy}},
  \emph{\bibinfo{title}{Finite-size Scaling}} (\bibinfo{publisher}{Elsevier},
  \bibinfo{address}{Amsterdam}, \bibinfo{year}{1988}).

\bibitem[{\citenamefont{Melchert}(2009)}]{autoscale}
\bibinfo{author}{\bibfnamefont{O.}~\bibnamefont{Melchert}}
  (\bibinfo{year}{2009}), \eprint{arXiv:0910.5403}.

\bibitem[{\citenamefont{Melchert and Hartmann}(2007)}]{melchert2007}
\bibinfo{author}{\bibfnamefont{O.}~\bibnamefont{Melchert}} \bibnamefont{and}
  \bibinfo{author}{\bibfnamefont{A.~K.} \bibnamefont{Hartmann}},
  \bibinfo{journal}{Phys. Rev. B} \textbf{\bibinfo{volume}{76}},
  \bibinfo{pages}{174411} (\bibinfo{year}{2007}).

\bibitem[{\citenamefont{Melchert and Hartmann}(2009)}]{melchert2009}
\bibinfo{author}{\bibfnamefont{O.}~\bibnamefont{Melchert}} \bibnamefont{and}
  \bibinfo{author}{\bibfnamefont{A.~K.} \bibnamefont{Hartmann}},
  \bibinfo{journal}{Phys. Rev. B} \textbf{\bibinfo{volume}{79}},
  \bibinfo{pages}{184402} (\bibinfo{year}{2009}).

\end{thebibliography}

\end{document}